\documentclass[sort&compress,final]{aipproc}\layoutstyle{6x9}
\usepackage{amsmath}\usepackage{amssymb}\usepackage{booktabs}
\begin{document}\title{Bethe--Salpeter Equations with Instantaneous
Confinement: Establishing Stability of Bound States}
\author{Wolfgang LUCHA}{address={Institute for High
Energy Physics, Austrian Academy of Sciences,\\Nikolsdorfergasse
18, A-1050 Vienna, Austria}}\begin{abstract}Analytic scrutinies of
Salpeter equations with confining interactions may identify
kernels that describe bound states free from the notorious
instabilities encountered in numerical evaluations.\end{abstract}

\maketitle

\section{Too na\"ively implemented confinement might allow for unstable
bound states}The perhaps most popular three-dimensional reduction
of the Bethe--Salpeter framework for the description of bound
states within quantum field theories is the {\em Salpeter
equation\/}, found in the {\em instantaneous\/} limit of the
Bethe--Salpeter formalism if assuming in addition {\em free\/}
propagation of the bound-state constituents. Unfortunately,
depending on the chosen {\em Dirac structure\/} of the
Bethe--Salpeter kernel (which encodes all interactions between the
bound-state constituents) even for Salpeter equations with {\em
confining\/} interactions, arising, e.g., from quantum
chromodynamics, the predicted --- supposedly stable --- bound
states exhibit instabilities, probably of similar nature as those
observed in Klein's paradox. The observation of this kind of
instability has been reported by numerous but to a large extent
{\em numerical\/} studies investigating confining interaction
potentials of in configuration space either strictly linear shape
or a form interpolating between harmonic-oscillator and linear
behaviour \cite{StabSal,Kopaleishvili}. Generally, for {\em
linearly\/} confining Bethe--Salpeter kernels being mixtures of
Lorentz scalar plus time-component Lorentz vector, stability can
be assured only if the Dirac structure of this kernel is
predominantly of time-component Lorentz-vector~nature.

Building on experience acquired in previous analyses, focused to
the simpler ``{\em reduced\/} Salpeter equation''
\cite{Lucha07:HORSE,Lucha07:StabOSS-QCD@Work07} and its
improvement \cite{Lucha05:IBSEWEP(a),Lucha05:IBSEWEP(b)} by
including dressed propagators for all bound-state constituents
\cite{Lucha05:EQPIBSE, Lucha07:SSSECI-Hadron07,Lucha07:HORSEWEP},
this investigation aims at rigorous analytic proofs
\cite{Lucha08-C8} of the stability of solutions of the (now) {\em
full\/} Salpeter equation with confining interactions (such as
harmonic-oscillator potentials) of frequently used Lorentz
structures, in order to identify all those kernels for which
bound-state stability can be taken as granted {\em ab initio\/}.
We regard bound states as stable if their energy (or, in their
center-of-momentum system, mass) eigenvalues belong to a {\em
real\/}, {\em discrete\/} portion of the corresponding {\em
spectrum\/} that is {\em bounded from below\/}. Such discussion
might provide further insight into the reasons why, for Lorentz
structures different from a time-component Lorentz vector,
instabilities~arise.

\section{Full (in contrast to reduced) Salpeter formalism for
relativistic {\em fermion--antifermion\/} bound states}Assuming,
as usual, the Lorentz structures of the effective couplings of
both fermion and antifermion to be represented by {\em
identical\/} Dirac matrices (generically called $\Gamma$
hereafter) and denoting the associated Lorentz-scalar interaction
function by $V_\Gamma(\mathbf{p},\mathbf{q}),$ our {\em Salpeter
eigenvalue equation\/} governing a chosen fermion--antifermion
bound state of mass $M$ and distribution $\Phi(\mathbf{p})$ of
internal momenta $\mathbf{p}$ (its ``Salpeter amplitude'') reads
in the rest~frame\begin{align}\Phi(\mathbf{p})
=\int\frac{\mathrm{d}^3q}{(2\pi)^3}\,\sum_\Gamma
V_\Gamma(\mathbf{p},\mathbf{q})&
\left(\frac{\Lambda^+(\mathbf{p})\,\gamma_0\,\Gamma\,\Phi(\mathbf{q})\,
\Gamma\,\Lambda^-(\mathbf{p})\,\gamma_0}{M-2\,E(p)}\right.\nonumber\\[.42ex]
&\hspace{-.3ex}\left.-\frac{\Lambda^-(\mathbf{p})\,\gamma_0\,\Gamma\,
\Phi(\mathbf{q})\,\Gamma\,\Lambda^+(\mathbf{p})\,\gamma_0}{M+2\,E(p)}
\right),\label{Eq:SEE}\end{align}with one-particle kinetic energy
$E(p)$ and energy projection operators $\Lambda^\pm(\mathbf{p}),$
defined~by$$E(p)\equiv\sqrt{p^2+m^2}\ ,\qquad p\equiv|\mathbf{p}|\
,\qquad\mbox{and}\qquad\Lambda^\pm(\mathbf{p})\equiv\frac{E(p)\pm
\gamma_0\,(\mathbf{\gamma}\cdot\mathbf{p}+m)}{2\,E(p)}\ ;$$here
$m$ labels the common mass of the bound fermion and the associated
antiparticle. The projector structure of Eq.~(\ref{Eq:SEE})
constrains all its solutions, the Salpeter
amplitudes~$\Phi(\mathbf{p}),$~to\begin{equation}
\Lambda^\pm(\mathbf{p})\,\Phi(\mathbf{p})\,\Lambda^\pm(-\mathbf{p})=0\
.\label{Eq:POSC}\end{equation}

\section{General features of the eigenvalue spectra of full
Salpeter equations}The structural equivalence of Salpeter's
equation (\ref{Eq:SEE}) to the random-phase-approximation equation
\cite{Resag94}, familiar from the investigation of collective
excitations in nuclear physics, or direct inspection allow one to
identify various characteristics common to all solutions:
\begin{itemize}\item The random-phase-approximation structure of the
Salpeter equation guarantees that the {\em squares\/} $M^2$ of the
mass eigenvalues are real. In general, the spectrum itself is {\em
not\/} necessarily real and, even where it is proven to be real,
it is {\em not\/} bounded~from~below.\item The most important
applications of the {\em instantaneous\/} Bethe--Salpeter
formalism are those which adopt interaction kernels only composed
of momentum-space potential functions
$V_\Gamma(\mathbf{p},\mathbf{q})$ and Dirac couplings $\Gamma$
that satisfy $V^\ast_\Gamma(\mathbf{p},\mathbf{q})=
V_\Gamma(\mathbf{p},\mathbf{q})= V_\Gamma(\mathbf{q},\mathbf{p})$
and $\gamma_0\,\Gamma^\dag\,\gamma_0=\pm\Gamma.$ Their spectra of
mass eigenvalues $M$ in the complex-$M$ plane are just unions of
{\em real\/} opposite-sign pairs $(M,-M)$ and/or {\em imaginary\/}
points $M=-M^\ast$.\end{itemize}Since eigenvalues embedded in a
continuous part of the spectrum may cause instabilities, the
nature of the {\em entire\/} spectrum proves to be crucial: Apart
from demanding eigenvalues to be {\em real\/} and {\em bounded
from below\/} the bound-state stability sought is established if
either eigenvalues and continuous spectrum are {\em disjoint\/} or
the spectrum is {\em purely discrete\/} at~all.

As a consequence of the constraint (\ref{Eq:POSC}), any Salpeter
amplitude $\Phi(\mathbf{p})$ may be expanded in terms of at most
eight independent components called, say, $\phi_i(\mathbf{p}),$
$i=1,\dots,8.$ We start this quest for bound-state stability at a
system requiring the {\em least\/} number of components.

\section{Harmonic-oscillator confinement simplifies the Salpeter
{\em integral\/} equation to an easier-to-handle {\em system\/} of
{\em radial\/} eigenvalue {\em differential equations\/}}All the
instabilities under consideration are expected to show up first in
the {\em pseudoscalar\/} sector \cite{StabSal}. Consequently, the
primary targets of all analyses of the present kind are bound
states with {\em spin-parity-charge conjugation assignment\/}
$J^{PC}=0^{-+}.$ A Salpeter amplitude $\Phi(\mathbf{p})$
describing such a state involves just {\em two independent
components\/} $\phi_1(\mathbf{p})$ and $\phi_2(\mathbf{p})$:
$$\Phi(\mathbf{p})=\left[\phi_1(\mathbf{p})\,
\frac{\gamma_0\,(\mathbf{\gamma}\cdot\mathbf{p}+m)}{E(p)}
+\phi_2(\mathbf{p})\right]\gamma_5$$is the unique form of any
Salpeter amplitude $\Phi(\mathbf{p})$ for a fermion--antifermion
bound state of total spin $J,$ parity $P=(-1)^{J+1},$ and
charge-conjugation quantum number $C=(-1)^J.$

Let our interaction kernel be of convolution type
[$V_\Gamma(\mathbf{p},\mathbf{q})
=V_\Gamma(\mathbf{p}-\mathbf{q})$], arising from a central
potential $V(r),$ where $r\equiv|\mathbf{x}|,$ in configuration
space. Using a harmonic-oscillator potential $V(r)=a\,r^2,$ where
$a\ne0$ avoids triviality, to {\em exemplify\/} our line of
reasoning, the Salpeter equation (1) reduces to a set of
second-order differential equations utilizing~only a single
differential operator (which is just the Laplacian
$\Delta\equiv\mathbf{\nabla}\cdot\mathbf{\nabla}$, acting on
$\ell=0$ states)$$D\equiv\frac{\mathrm{d}^2}{\mathrm{d}p^2}
+\frac{2}{p}\,\frac{\mathrm{d}}{\mathrm{d}p}\ .$$Assuming massless
bound-state constituents, $m=0,$ facilitates our study of this
problem.

In our analysis interaction kernels of time-component
Lorentz-vector, $\Gamma\otimes\Gamma=\gamma^0\otimes\gamma^0$, and
Lorentz-scalar, $\Gamma\otimes\Gamma=1\otimes1$, nature may be
discussed simultaneously if
introducing$$\sigma=\left\{\begin{array}{rll}+1&\qquad\mbox{for
$\Gamma\otimes\Gamma=\gamma^0\otimes\gamma^0$}&\mbox{(time-component
Lorentz-vector interactions)}\\[.21ex]-1&\qquad\mbox{for
$\Gamma\otimes\Gamma=1\otimes1$}&\mbox{(Lorentz-scalar
interactions)}\end{array}\right.$$as a discriminating parameter.
By factorizing off all dependence on angular variables,~the {\em
Salpeter equation\/} for harmonic-oscillator interactions of {\em
time-component Lorentz-vector\/} or {\em Lorentz-scalar\/} Dirac
structure reduces to a system of two radial {\em differential\/}
equations:
\begin{align}\left(2\,p-a\,\sigma\,D\right)\phi_2(p)&=M\,\phi_1(p)\
,\nonumber\\[.22ex]
\left[2\,p-a\left(D-\frac{2}{p^2}\right)\right]\phi_1(p)&=M\,\phi_2(p)\
.\label{Eq:RSE}\end{align}Adding the sets for $\sigma=+1$ and
$\sigma=-1$, we find the Salpeter equation for {\em Lorentz-scalar
plus time-component Lorentz-vector mixing\/},
$\Gamma\otimes\Gamma=\xi\,\gamma^0\otimes\gamma^0
+\eta\,1\otimes1$, where~$\xi,\eta\in{\mathbb R}$:
\begin{align}\left[2\,p-a\,(\xi-\eta)\,D\right]\phi_2(p)&
=M\,\phi_1(p)\ ,\nonumber\\[.22ex]
\left[2\,p-a\,(\xi+\eta)\left(D-\frac{2}{p^2}\right)\right]\phi_1(p)&
=M\,\phi_2(p)\ .\label{Eq:RSE-STVM}\end{align}The units of the
momentum $p$ can be chosen, without loss of generality, such that
$|a|=1.$

However, in spite of the considerable technical simplification
achieved by reduction of the Salpeter integral equation to a set
of differential equations, for the kernels of different Lorentz
structure we shall perform the spectral analyses still due on a
case-by-case basis.

\section{Confining interaction kernels of time-component
Lorentz-vector Dirac structure
($\Gamma\otimes\Gamma=\gamma^0\otimes\gamma^0$)} For the
time-component Lorentz-vector Dirac couplings ($\sigma=+1$), the
mere existence of bound states requires the coupling $a$ to be
positive: $a>0.$ In this case, an entirely {\em analytic\/} proof
of the (so far basically numerically established) stability of the
bound~states may be constructed. Expressed by the positive
self-adjoint operators on the Hilbert space~$L^2({\mathbb R}^3)$
$$A\equiv-\Delta+2\,r=A^\dag\ge0\ ,\qquad
B\equiv-\Delta+2\,r+\frac{2}{r^2}=B^\dag\ge0\ ,\qquad
r\equiv|\mathbf{x}|\ ,$$our radial Salpeter equation
(\ref{Eq:RSE}) with harmonic-oscillator confining interactions is
clearly equivalent to the (vanishing angular momentum sector of
the) matrix eigenvalue problem
$$\begin{array}{l}A\,f_2=M\,f_1\\B\,f_1=M\,f_2\end{array}
\qquad\Longleftrightarrow\qquad
\left(\begin{array}{cc}0&A\\B&0\end{array}\right)
\left(\begin{array}{c}f_1\\f_2\end{array}\right)=M
\left(\begin{array}{c}f_1\\f_2\end{array}\right),\qquad f_1,f_2\in
L^2({\mathbb R}^3)\ .$$Reiteration of the L.H.S.\ operation yields
an equivalent problem for {\em real\/} eigenvalues~$M^2$:
$$\left(\begin{array}{cc}A\,B&0\\0&B\,A\end{array}\right)
\left(\begin{array}{c}f_1\\f_2\end{array}\right)=M^2
\left(\begin{array}{c}f_1\\f_2\end{array}\right)
\qquad\Longleftrightarrow\qquad
\begin{array}{l}A\,B\,f_1=M^2\,f_1\ ,\\B\,A\,f_2=M^2\,f_2\
.\end{array}$$

The spectral theorem for self-adjoint operators allows one to
define the unique positive self-adjoint square root of $A$:
$A^{1/2}=(A^{1/2})^\dag\ge0.$ Setting $g\equiv A^{1/2}\,f_2$
converts the relation $B\,A\,f_2=M^2\,f_2$ to an eigenvalue
equation, $Q\,g=M^2\,g,$ of the {\em positive self-adjoint\/}
operator $Q\equiv A^{1/2}\,B\,A^{1/2}=Q^\dag\ge0.$ Obviously, our
operators $A$ and $B$ satisfy the {\em inequality\/} $A\le B;$
left and right multiplication by $A^{1/2}$ transforms this
inequality into $A^2\le A^{1/2}\,B\,A^{1/2}\equiv Q.$ A {\em
theorem\/} about the spectrum of a Hamiltonian with potential
increasing beyond bounds claims that a Schr\"odinger operator
$H\equiv-\Delta+V,$ defined as sum of quadratic forms, with
locally bounded, positive, infinitely rising potential $V$
[$V(\mathbf{x})\to+\infty$ for $r\to\infty$] has {\em purely
discrete\/} spectrum; application of this theorem guarantees that
the spectrum of $A$ is {\em purely discrete\/}. The spectral
theorem for self-adjoint operators allows for a representation of
$A^2$ in terms of the same projection-valued spectral measure as
$A;$ consequently, the spectrum of $A^2$ is also {\em entirely
discrete\/}. Trivially, the positive operators $A^2$ and $Q$ are
bounded from below. Combining
\cite{MMP-OI(a),MMP-OI(b),MMP-OI(c),MMP-OI(d),MMP-OI(e),MMP-OI(f)}
the characterization of all {\em discrete\/} eigenvalues of an
arbitrary self-adjoint operator bounded from below by the
minimum--maximum principle with the operator inequality $A^2\le Q$
shows that the spectrum of our real squared mass eigenvalues $M^2$
of $Q$ must be {\em purely discrete\/} and therefore the spectrum
of bound-state masses $M$~too.

This stability of {\em full\/}-Salpeter bound states, proven for
time-component Lorentz-vector harmonic-oscillator interaction with
positive strength $a>0,$ {\em excludes\/} the possibility that, in
the case $\sigma=+1,$ {\em instabilities\/} of bound states are
induced by energy eigenvalues being embedded in a continuous
spectrum of the Salpeter operator controlling the bound states,
instead of belonging to its discrete spectrum, as expected for
true confinement. The proof cannot be transferred to $a<0$ as in
this case the counterparts of our operators $A$ and $B$~are
$$\widetilde A\equiv\Delta+2\,r\ ,\qquad\widetilde
B\equiv\Delta+2\,r-\frac{2}{r^2}\ .$$For both operators
$\widetilde A$ and $\widetilde B$, positivity is lost. This fact
invalidates most steps~of~our~proof.

\section{Confining interaction kernels of Lorentz-scalar
($\Gamma\otimes\Gamma=1\otimes1$) or Lorentz-pseudoscalar
($\Gamma\otimes\Gamma=\gamma_5\otimes\gamma_5$) Dirac structure}In
the limit of the bound-state constituents being exactly {\em
massless\/}, the Salpeter equations for {\em Lorentz-scalar\/}
[i.e., $\sigma=-1$ in Eq.~(\ref{Eq:RSE})] and {\em
Lorentz-pseudoscalar\/} interaction kernels (must) become
identical, as the anticipated manifestation of chiral symmetry.
The matrix eigenvalue problem is formulated in terms of the
operator pairs ($\widetilde A,B$) for $a>0$ and ($A,\widetilde B$)
for $a<0.$ In both cases, one of these operators is {\em not
positive\/}, which spoils our~reasoning.

\section{Linear combinations of both time-component
Lorentz-vector ($\Gamma\otimes\Gamma=\gamma^0\otimes\gamma^0$) and
Lorentz-scalar ($\Gamma\otimes\Gamma=1\otimes1$) confining
interaction kernels}Introducing the abbreviations $\alpha\equiv
a\,\xi\in{\mathbb R},$ $\beta\equiv a\,\eta\in{\mathbb R},$ the
two operators entering in the Salpeter equation
(\ref{Eq:RSE-STVM}) for some mixture of scalar and time-component
vector kernels~read
$$\mathcal{A}\equiv(\alpha-\beta)\,(-\Delta)+2\,r\ ,\qquad
\mathcal{B}\equiv(\alpha+\beta)\left(-\Delta+\frac{2}{r^2}\right)+2\,r\
,\qquad r\equiv|\mathbf{x}|\ .$$Positivity of {\em both\/}
(symmetric) operators $\mathcal{A}$, $\mathcal{B}$ and presence of
{\em both\/} derivatives demands $\alpha-\beta>0$ {\em and\/}
$\alpha+\beta>0.$ These two relations restrain the couplings
$\alpha$ and $\beta$ to the range
\begin{equation}\alpha=|\alpha|>|\beta|\ge0\qquad\Longleftrightarrow\qquad
-1<\frac{\beta}{\alpha}<+1\quad\mbox{and}\quad\alpha>0\
.\label{Eq:S/TV}\end{equation}A proof similar to the above for a
pure time-component Lorentz-vector kernel establishes stability
{\em irrespective of the relative sign\/} of the two contributions
of unequal Lorentz type: the dominance of the time-component
Lorentz-vector kernel guarantees stability.~Table~\ref{Tab:1}
illustrates these findings for various parameter ratios
$\beta/\alpha$ within the tolerable region (\ref{Eq:S/TV}).

\begin{table}[hb]\caption{Lowest-lying positive
mass eigenvalues (in units of $\sqrt[3]{a}$) of the Salpeter
equation (\ref{Eq:RSE-STVM}) for several values $\beta/\alpha$ of
our couplings in the stability-compatible range
$-1\le\beta/\alpha<+1.$}
\begin{tabular}{lrrrrrrrrr}\toprule Level&\multicolumn{9}{c}
{$\displaystyle\frac{\beta}{\alpha}=\frac{\eta}{\xi}$}\\
\cmidrule{2-10}
&\multicolumn{1}{c}{$+0.99$}&\multicolumn{1}{c}{$+0.75$}
&\multicolumn{1}{c}{$+0.50$}&\multicolumn{1}{c}{$+0.25$}
&\multicolumn{1}{c}{$0.00$}&\multicolumn{1}{c}{$-0.25$}
&\multicolumn{1}{c}{$-0.50$}&\multicolumn{1}{c}{$-0.75$}
&\multicolumn{1}{c}{$-1.00$}\\[.4ex]\toprule
0&4.53&4.69&4.71&4.67&4.60&4.47&4.28&3.97&2.93\\
1&6.15&6.84&7.07&7.16&7.15&7.06&6.87&6.49&4.68\\
2&7.63&8.78&9.14&9.30&9.33&9.24&9.01&8.55&6.14\\
3&9.01&10.54&11.02&11.23&11.28&11.18&10.92&10.38&7.45\\
4&10.31&12.18&12.75&13.01&13.07&12.97&12.68&12.06&8.65\\
5&11.55&13.73&14.38&14.67&14.75&14.65&14.32&13.63&9.77\\
6&12.74&15.19&15.92&16.25&16.35&16.23&15.87&15.11&10.83\\
7&13.88&16.59&17.39&17.76&17.86&17.74&17.35&16.52&11.84\\
8&14.98&17.93&18.81&19.21&19.33&19.19&18.77&17.87&12.81\\
9&16.04&19.23&20.19&20.64&20.78&20.63&20.16&19.18&13.74\\
\bottomrule\end{tabular}\label{Tab:1}\end{table}

\section{Summary, Findings, Conclusions, and Perspectives}
Salpeter equations with potential functions rising to infinity in
configuration space do not automatically predict stable bound
states: for this to happen, the Lorentz behaviour of the involved
Bethe--Salpeter kernels too is crucial. Truly confining
interaction kernels can be singled out rather systematically by
requiring the emerging bound-state energy spectra to be both real
and discrete. This task is comparatively easy if the Salpeter
equation reduces to a differential equation (as for the harmonic
oscillator) but might be conducted~for~more general (say, linearly
confining) potentials if all integral-equation~issues can be
mastered.

\begin{theacknowledgments}I would like to express my honest
gratitude to Bernhard Baumgartner, Harald Grosse and Heide
Narnhofer for many interesting, stimulating, encouraging and
helpful discussions.\end{theacknowledgments}

\bibliographystyle{aipproc}\end{document}